\documentclass[draft]{agujournal2019}
\usepackage{url} 
\usepackage{lineno}
\usepackage{natbib}
\usepackage{amsmath}
\usepackage[inline]{trackchanges} 
\usepackage{soul}
\usepackage{xcolor}

\draftfalse

\journalname{AGU Advances}

\begin{document}

\title{A Lagrangian Perspective on the Growth of Midlatitude Storms}

\authors{Or Hadas\affil{1}, Yohai Kaspi\affil{1}}

\affiliation{1}{Department of Earth and Planetary Sciences, Weizmann Institute of Science, Rehovot, Israel}

\correspondingauthor{Or Hadas}{or.hadas@weizmann.ac.il}

\begin{keypoints}
\item To uncover the relation between the mean flow and storm growth, we analyze Lagrangianly all midlatitude storms from 83 years of ERA-5 data.
\item While the Lagrangian growth rate is found to increase monotonically with baroclinicity, the storm growth time exhibits an opposite trend.
\item We propose a quantitative non-linear relation between the local mean flow characteristics of the atmosphere and the intensity of storms.
\end{keypoints}

\begin{abstract}
Extratropical storms dominate midlatitude climate and weather and are known to grow baroclinicaly and decay barotropicaly. Traditionally, quantitative climatic measures of storm growth have been mostly based on Eulerian measures, taking into account the mean state of the atmosphere and how those affect eddy growth, but they do not consider the Lagrangian growth of the storms themselves. Here, using ERA-5 reanalysis data and tracking all extratropical storms (cyclones and anticyclones) from 83 years of data, we examine the actual growth of the storms and compare it to the Eulerian characteristics of the mean state as the storms develop. In the limit of weak baroclinicity, we find that baroclinicity provides a good measure for storm maximum intensity. However, this monotonic relationship breaks for high baroclinicity levels. We show that although the actual growth rate of individual storms monotonically increases with baroclinicity, the reduction in maximum intensity at high baroclinicity is caused by a decrease in storm growth time. Based on the Lagrangian analysis, we suggest a nonlinear correction to the traditional linear connection between baroclinicity and storms' activity. Then, we show that a simplified model of storm growth, incorporating the baroclinicity effect on the vertical tilt of anomalies, reproduces the observed nonlinear relationship. Expanding the analysis to include the mean flow's barotropic properties highlights their marginal effect on storm growth rate, but the crucial impact on growth time. Our results emphasize the potential of Lagrangianly studying storm dynamics to advance understanding of the midlatitude climate.
\end{abstract}

\section*{Plain Language Summary}
The midlatitude climate is shaped by storms. Their growth is primarily driven by baroclinic instability, a process converting vertical wind shear into storms' energy. Understanding how climatic changes impact this growth is essential for advancing midlatitude climate and weather understanding. In this study, we systematically evaluated the growth of storms using ERA-5 reanalysis data and tracks of all storms over 83 years. The growth of individual storms is compared to baroclinicity, a widely used measure of storm activity that is primarily based on vertical wind shear. Our findings indicate that storm intensity increases with baroclinicity under mild conditions, this relationship breaks down under extreme conditions. While the growth rate of storms continues to increase with baroclinicity, this shift is due to a decrease in the storm growth time with baroclinicity. As a result, a nonlinear correction to the traditional linear relation between baroclinicity and storm activity is proposed. Expanding the analysis to include the horizontal wind shear emphasizes its minimal impact on storm growth rate but underscores its crucial influence on growth time. These results highlight the importance of systematic study of large quantities of storms.

\section{Introduction}

\begin{figure*}[t] 
\begin{centering}
\includegraphics[width=15pc]{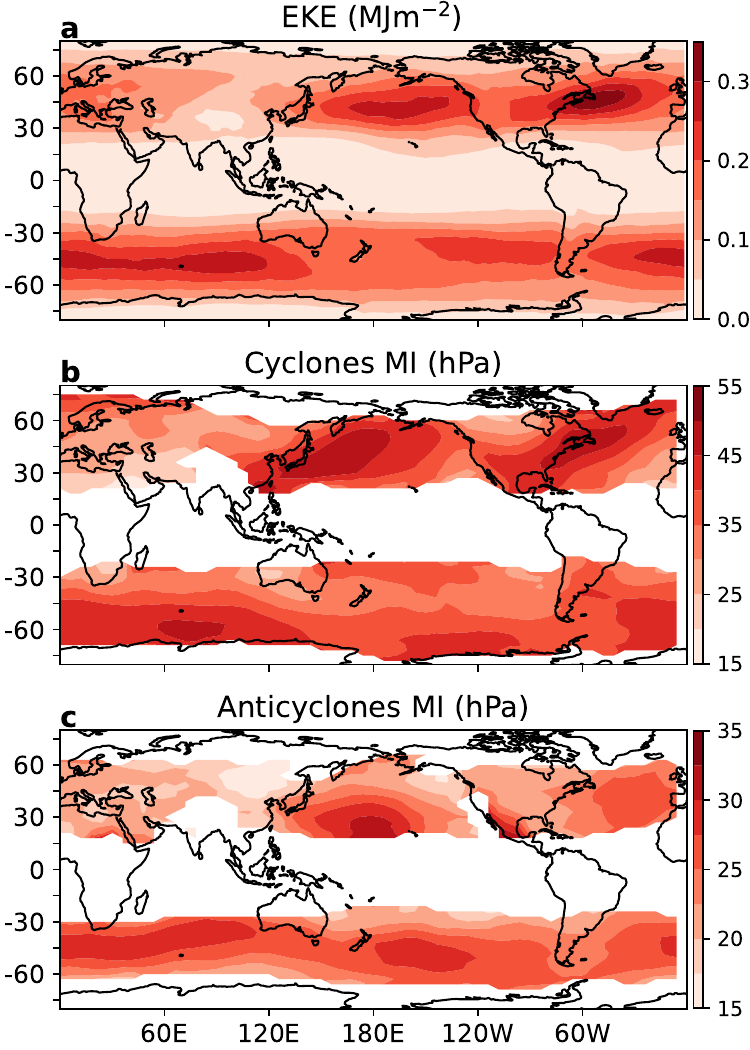} \par\end{centering} \caption{\label{fig:growthtime_map}Climatolagy of vertically integrated Eddy Kinetic Energy (EKE, MJm$^{-2}$, a) and Maximum Intensity (MI, hPa, as defined in Sec.~2.2) of cyclones (b) and anticyclones (c) during winter (December-January-February for the NH and June-July-August for the SH). EKE $\equiv \frac{1}{2g}\int_{850}^{300}(u'^2 + v'^2)dp$, where $p$ is the pressure, $g$ is the gravitational acceleration, $u'$ and $v'$ are the zonal and meridional wind anomalies filtered for 2-10 days and zonal wavenumbers 4-20. The MI of each storm is counted in all grid boxes it passes from genesis to MI. The MI maps' resolution is 6$^\circ\times$3$^\circ$, and they are smoothed using a convolution with 3$\times$3 filter. Only grid points averaging more than 200 storms are included. Winter climatology is shown due to its strong spatial variability. These results are consistent with previous findings \citep[e.g.,][]{hoskins2002, hoskins2005}.}
\end{figure*}

The midlatitude momentum, heat, and moisture fluxes display significant spatial variability, primarily concentrated in regions referred to as the \textquotedblleft storm tracks\textquotedblright{} \citep{Chang2002,hoskins2002,hoskins2005}. Three primary storm tracks have been identified: the Northern Hemisphere (NH) Atlantic and Pacific storm tracks and the Southern Hemisphere (SH) storm track. Moreover, the midlatitude climate displays a significant temporal variability, which is primarily seasonal \citep{nakamura2004seasonal,Hoskins2019}. From an Eulerian perspective, these variations are typically quantified using climatological metrics such as Eddy Kinetic Energy (EKE, Fig.~\ref{fig:growthtime_map}a), which provide insights into the large-scale statistical properties of storm tracks. From a Lagrangian perspective, these spatial and temporal variabilities arise from substantial differences in cyclonic and anticyclonic activity among various parts of Earth (Fig.~\ref{fig:growthtime_map}b,c), serving as the main driver for the weather variability throughout the midlatitudes \citep{Yau2020}.

These variations in the storm tracks have been extensively studied, often analyzed through a three-step approach \citep[e.g.,][]{Shaw2016}. First, atmospheric flow is decomposed (via spatial or temporal filtering) into the mean flow, representing large-scale, slowly varying features like jet streams and eddy fields dominated by cyclonic and anticyclonic activity. Second, the effect of atmospheric forcing changes on the mean state is assessed, leveraging the simplification achieved through averaging. Third, the storm track response is inferred from the eddy state’s reaction to mean-state changes. However, due to the nonlinear complexity of eddy fields, exact analytical solutions are unavailable, and numerical models depend heavily on specific settings. Despite these challenges, theoretical frameworks linking mean and eddy flow have been developed, including baroclinic instability models \citep[e.g.,][]{charney1947,Eady1949,Phillips1954}, potential vorticity tendency decomposition \citep[e.g.,][]{Hoskins1985,Davis1991,Tamarin2017mechanisms}, and energetics analyses \citep[e.g.,][]{Lorenz1955,Peixoto1974,Okajima2022}.

These methods are typically applied from an Eulerian perspective, focusing on how changes in properties of the mean flow drive climatological changes in storm tracks activity. For example, \citet{Lehmann2014future} showed a strong connection between baroclinicity and EKE in CMIP5 models, and \citet{Kang2024anthropogenic} demonstrated that anthropogenic aerosols significantly alter mean baroclinicity, explaining much of the NH summertime circulation weakening. Beyond baroclinicity, \citet{Chemke2022} attributed the recent intensification of SH EKE—unaccounted for in CMIP6 models—to trends in the barotropic properties of the mean flow.

While the Eulerian approach has greatly advanced our understanding of midlatitude climate, it has key limitations. Averaging mean and eddy flow properties (e.g., energy conversion terms in Sec.~\ref{subsec:Energetic-Perspective}) make it difficult to isolate individual contributions, limiting causal inference \citep{Vallis2017atmospheric}. It also blends the impacts of different weather systems (cyclones, anticyclones, atmospheric rivers), which have distinct dynamics. For example, comparing the climatology of EKE (Fig.~\ref{fig:growthtime_map}a) to the climatology of the Maximum Intensity of cyclones and anticyclones (MI, Fig.~\ref{fig:growthtime_map}b,c), reveals that cyclones tend to be strongest poleward and upstream of peak EKE, while anticyclones follow the opposite trend.

An example of a key limitation of the Eulerian perspective is the nonlinear relationship between baroclinicity and EKE. During midwinter over the Pacific, storm activity decreases despite maximum baroclinicity \citep{Nakamura1992}, a puzzle attributed to several mechanisms \citep[e.g.,][]{Nakamura2002,Harnik2004,schemm2019efficiency}. Additionally, \citet{Haualand2021relative} found that a sharper tropopause can suppress baroclinic instability, indicating that a stronger jet does not necessarily translate to enhanced instability.

Recently, studies have addressed some of these limitations by adopting a Lagrangian perspective—tracking individual storms through space and time—to reveal key atmospheric mechanisms, improve causal inference, and better capture actual weather patterns \citep[e.g.,][]{Tamarin-Brodsky2017,Tamarin2019,Schemm2021,Hadas2021suppression,Tsopouridis2021,Kang2021b,Okajima2021,Okajima2022,Hadas2023,Okajima2024}. From this perspective, the eddy mean flow interaction could be expressed in terms of how a slowly evolving climate (longer than 30 days) influences the dynamics of synoptic-scale storm activity (about one week). Although any single storm makes only a minor contribution to the mean flow, the collective impact of many storms is substantial. Therefore, understanding how the mean flow affects each storm is a crucial first step toward a comprehensive Lagrangian framework for eddy–mean flow interactions, which is vital for understanding midlatitude climate.

Previous Lagrangian-based research on storm responses to the mean flow has used three main approaches: case studies \citep[e.g.,][]{Orlanski1991,Riviere2006}, which capture real-world processes but are limited in scope; idealized simulations \citep[e.g.,][]{Orlanski1993,Riviere2013,Hadas2021suppression}, which allow for systematic investigation, but may not fully represent real conditions; and region-specific analyses \citep[e.g.,][]{Schemm2021,Tsopouridis2021,Okajima2023}, whose findings may be tied to local configurations. 

To address these limitations, this study systematically examines how storm growth is influenced by mean flow properties commonly used in Eulerian studies. The analysis is based on 83 years of ERA-5 data (Sec.~\ref{subsec:Reanalysis-data}) and tracks extratropical storms across all seasons and regions (Sec.~\ref{subsec:Storm-tracking}), providing a vast dataset of approximately 100,000 cyclones and 50,000 anticyclones. The properties of the mean flow are characterized using the energetic perspective (Sec.~\ref{subsec:Energetic-Perspective}). The analysis begins by examining the response of storm growth to the baroclinic characteristics of the mean flow (Sec.{~\ref{subsec:Correlations}}). These trends are then interpreted using an idealized storm growth model (Sec.~\ref{subsec:theory}). Finally, the analysis is extended to include the barotropic characteristics of the flow (Sec.{~\ref{subsec:dist})}. A brief summary of the results is provided in Sec.~\ref{sec:summary}.

\section{Methods\label{sec:Methods}}

\subsection{Reanalysis data\label{subsec:Reanalysis-data}}

Data from the European Center for Medium-Range Weather Forecasts ERA-5 reanalysis \citep{Hersbach2020} between 1940 and 2023 is used to assess the current climate. The ERA-5 estimates atmospheric variables at a horizontal resolution of 31 km and 137 vertical levels. The three-hour Sea Level Pressure (SLP) is used to identify the tracks of cyclones and anticyclones. Three hourly temperature, geopotential, and horizontal wind data are used to characterize the mean flow. The data is down-sampled to 1.5$^{\circ}\times$1.5$^{\circ}$.

\subsection{Storm tracking and composites\label{subsec:Storm-tracking}}

A feature point tracking algorithm \citep{hodges_1995,Tamarin2016a} is applied to SLP data to identify and characterize extratropical cyclones and anticyclones. The data is smoothed to T63 resolution to reduce noise, and zonal wavenumbers 0–4 are removed to isolate synoptic-scale dynamics (using the methods described in \cite{hodges_1995}). The algorithm detects local SLP minima and maxima, tracks anomalies, and records storm position and magnitude. Storms are included if their magnitude increases for over 24 hours and they travel more than 1000 km. Additionally, storms peaking over terrain higher than 1 km are filtered out. To maximize data availability and to make the analysis as general as possible, storms from all over the world and from all seasons are included in the analysis. The final dataset includes approximately 100,000 cyclones and 50,000 anticyclones.

To quantify storm intensity, the pressure anomaly magnitude detected by the tracking algorithm is used. Since geostrophic balance implies that a given pressure anomaly corresponds to different wind strengths at different latitudes, storm intensity is defined as \citep[e.g.,][]{sanders1980synoptic}:

\begin{align}
    \text{Intensity}\equiv\frac{{\Delta}P}{\sin(\phi)},\label{eq:intensity}
\end{align}
where ${\Delta}P$ is the magnitude of the pressure anomaly identified by the tracking algorithm, and $\phi$ is its latitude.

\noindent Based on intensity, three measures characterize storm growth are constructed:
\begin{itemize}
    \item \textbf{Maximum Intensity} (MI, hPa) – the peak intensity reached along the track of the storm.
    \item \textbf{Growth Time} (GT, days) – the duration from genesis to maximum intensity.
    \item \textbf{Lagrangian Growth Rate} (LGR, hPa$\times$day$^{-1}$) – the average rate of intensification over the growth stage.
\end{itemize}
Notably, LGR does not have typical rate units (day$^{-1}$) because it is not normalized by the initial intensity, which is often small and highly sensitive to noise. To focus on significant events, the storms that had the 2\% lowest maximum intensity were removed.

To quantify the mean flow conditions experienced by storms during growth, the mean flow is averaged spatially within a composite box extending $\pm$2000 km zonally and $\pm$1000 km meridionally from the storm's SLP anomaly center and temporarily from identification to MI.

\subsection{Energetic perspective\label{subsec:Energetic-Perspective}}

In this section, we define mean flow quantities that capture its fundamental properties and can be used to analyze its influence on individual storm growth. To achieve this, we adopt the energetics framework, which directly incorporates the eddy-mean flow interactions, providing a clear physical interpretation of the constructed mean flow quantities. The energy exchange between the mean flow and eddies primarily occurs through two processes: baroclinic conversion, which transfers mean available potential energy to eddy available potential energy, and barotropic conversion, which transfers eddy kinetic energy to mean kinetic energy \citep{Lorenz1955}. Therefore, we seek to identify mean flow quantities that directly influence the rate of these energy conversions.

Starting from the equations for baroclinic and barotropic energy conversion \citep[][see \ref{sec:energy_der} for the derivation of this specific form]{Simmons1983barotropic,Orlanski1991}:
\begin{align}
C(P_{e},P_{m}) & =\underbrace{\frac{\rho g}{\overline{T}\overline{N}}\left(\overline{u'T'},\overline{v'T'}\right)}_{\text{Eddy part}}\cdot\underbrace{\frac{f}{\overline{N}}\left(\frac{\partial\overline{v}}{\partial z},\frac{\partial\overline{u}}{\partial z}\right)}_{\text{Mean part}}\label{Eq_CPEM},\\
C(K_{e},K_{m}) & =\underbrace{\rho\left(\overline{v'^{2}-u'^{2}},-\overline{u'v'}\right)}_{\text{Eddy part}}\cdot\underbrace{\left(\frac{\partial\overline{u}}{\partial x}-\overline{v}\frac{\tan\phi}{r_{e}},\frac{\partial\overline{u}}{\partial y}+\frac{\partial\overline{v}}{\partial x}+\overline{u}\frac{\tan\phi}{r_{e}}\right)}_{\text{Mean part}},\label{Eq_CKEM}
\end{align}
where $C(P_{e},P_{m})$ and $C(K_{e},K_{m})$ are the conversion between the eddy and mean potential and kinetic energy, respectively, $u$ and $v$ are the zonal and meridional components of the wind, respectively, $g$ is the gravitational constant, $f$ is the Coriolis parameter, $\rho$ is the density of air, $T$ is temperature, $r_{e}$ is the radius of Earth and $\overline{N}$ is the Brunt-Väisälä frequency:

\begin{align}
\overline{N}=\sqrt{-\frac{{\rho}g^2}{{\overline{\theta}}}\frac{{\partial}\overline{\theta}}{{\partial}p}}, \label{eq:bvf}
\end{align}
where $p$ is the pressure, $\theta$ is the potential temperature. Overbar and prime represent mean and deviation from a mean. In this study, the mean is defined as a low pass from 30 days, which is much longer than the timescale of the storms (about 2-7 days). Additionally, to focus on large-scale patterns, a lowpass from wavenumber 5 was used. Eddy terms are not studied in this research directly, only through the Lagrangian growh properties. 
 
Next, the mean flow quantities used for regression against individual storm growth are constructed based on the mean-dependent component of the conversion rates (right brace in Eqs.~\ref{Eq_CPEM} \& \ref{Eq_CKEM}).

 \begin{align}
     \left(\widehat{\text{CP}}_{x},\widehat{\text{CP}}_{y}\right)&=\frac{f}{\overline{N}}\left(\frac{\partial\overline{v}}{\partial z},\frac{\partial\overline{u}}{\partial z}\right),\\ \left(\widehat{\text{CK}}_{x},\widehat{\text{CK}}_{y}\right)&=\left(\frac{\partial\overline{u}}{\partial x}-\overline{v}\frac{\tan\phi}{r_{e}},\frac{\partial\overline{u}}{\partial y}+\frac{\partial\overline{v}}{\partial x}+\overline{u}\frac{\tan\phi}{r_{e}}\right),
 \end{align}
 These mean components are measured in rate units (day$^{-1}$) and will therefore be referred to in the plural as shear rates. The baroclinic shear rates are mainly proportional to the vertical shear of the winds, while barotropic shear rates are mainly proportional to the horizontal shear of the winds. The shear rates can be interpreted as the barotropic and baroclinic energy conversion rates for a given storm energy, as the eddy part is proportional to the anomalies squared, similar to EKE.

To simplify the baroclinic components of the shear rate, constant stratification, and wind shear are assumed, eliminating variability in the vertical dimension. Additionally, the mean wind component is assumed to vanish at the surface, as surface winds are generally weak and poorly captured by the ERA-5 reanalysis \citep{Belmonte2019}. The resulting estimate for the baroclinic part is:

\begin{equation} 
\left(\widehat{\text{CP}}_{x},\widehat{\text{CP}}_{y}\right)=\frac{f}{NH_{300}}\left(\overline{v}_{300},\overline{u}_{300}\right)\label{eq:CP},
\end{equation} 
where subscript denotes the pressure level that is used. $N$ is estimated as:

\begin{align}
N=\sqrt{-\frac{\rho_{500}g^{2}}{\overline{\theta}_{500}}\frac{\overline{\theta}_{300}-\overline{\theta}_{850}}{p_{300}-p_{850}}},
\end{align}
where $\rho$ is calculated based on the ideal gas law. The magnitude of the baroclinic shear rate vector is used to define the baroclinicity:

\begin{align}
\text{BC}\equiv\frac{f}{NH_{300}}\sqrt{\overline{v}_{300}^2+\overline{u}_{300}^2}.\label{eq:EGR} 
\end{align}
This definition of baroclinicity is closely related to the Eady Growth Rate \citep{Eady1949}, which represents the linear baroclinic instability prediction for the growth rate of anomalies and is widely used in Eulerian studies to analyze storm activity responses \citep[e.g.,][]{Lehmann2014future}.

To simplify the barotropic shear rates, we consider that barotropic conversion primarily occurs in the upper atmosphere \citep{Peixoto1992}. Therefore, the mean flow at 300 hPa is used to assess the barotropic component of the shear rate:

\begin{equation} \left(\widehat{\text{CK}}_{x},\widehat{\text{CK}}_{y}\right)=\left(\frac{\partial\overline{u}_{300}}{\partial x}-\overline{v}_{300}\frac{\tan\phi}{r_{e}},\frac{\partial\overline{u}_{300}}{\partial y}+\frac{\partial\overline{v}_{300}}{\partial x}+\overline{u}_{300}\frac{\tan\phi}{r_{e}}\right),\label{eq:CK} \end{equation} 
where $\widehat{\text{CK}}_{x}$ and $\widehat{\text{CK}}_{y}$ are known as stretching deformation and shearing deformation, respectively \citep{Mak1989}.

For individual storms, additional energy sources may play a significant role, such as ageostrophic energy fluxes \citep[e.g.,][]{Orlanski1991} and diabatic heating \citep[e.g.,][]{Weijenborg2020diabatic}. These sources could indirectly influence our analysis. For example, diabatic heating is crucial in maintaining the baroclinic structure of the atmosphere \citep{Papritz2015}. Therefore, future work should aim to quantify the specific physical mechanisms underlying the observed relationships, including the contribution of diabatic heating.

\section{The Lagrangian response to baroclinicity}

\subsection{The empirical relation between growth and baroclinicity\label{subsec:Correlations}}

\begin{figure}[t]
\begin{centering}
\includegraphics[width=31pc]{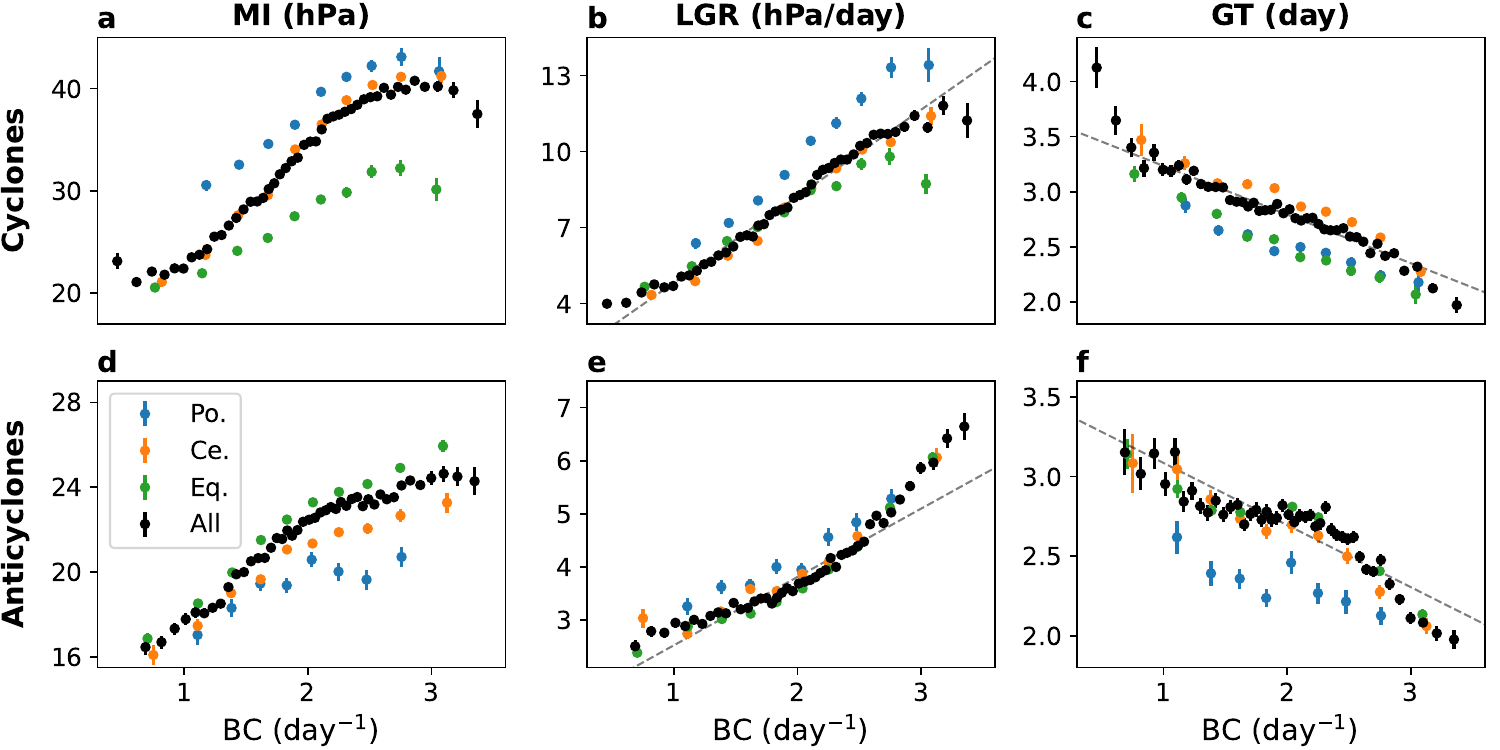}
\par\end{centering}
\centering{}
\caption{\label{fig:prop_corr}The relation between cyclones' MI (hPa, a), LGR (hPa$\cdot$day$^{-1}$, b) and GT (day, c) to the composite mean BC (day$^{-1}$) for all cyclones (black), cyclones poleward of the jet core (blue), around the center of the jet core (orange) and equatorward of the jet core (green). The dashed curve is the linear regression between the BC and the Lagrangian properties. The parameters of the fit appear in Tab.~S1. (d-f) Same as (a-c), but for anticyclones. Fitted parameters for anticyclones appear in Tab.~S2. Error bars represent the standard error within the BC clusters. Only clusters that have more than 100 storms are plotted. Poleward, around the center, and equatorward of the jet are defined as cyclones (anticyclones) that experienced meridional shear of zonal wind lower than -0.75 (-0.5), between -0.1 (-0.15) and 0.1 (0.15) and higher than 0.75 (0.5) (day$^{-1}$), respectively. Wider margins of shear are used for anticyclones because they are fewer. The black dots reach higher BC values as including storms from all parts of the jet increases the maximum BC at which 100 storms are obtained.}
\end{figure}

The study first examines how storm growth responds to the baroclinic characteristics of the mean flow (baroclinicity), a key driver of storm development. This is achieved by calculating the baroclinicity (BC, Eq.~\ref{eq:EGR}) experienced by storms, averaged both spatially within a composite box and temporally from genesis to MI (Sec.~{\ref{subsec:Storm-tracking}}). Storms are then classified based on their average BC using a K-means algorithm \citep{Scikit-learn}, which optimally partitions them into clusters. The average growth properties (e.g., MI) are then plotted against the mean BC for each cluster. By averaging storms that experienced similar BC, the analysis statistically removes variability due to uncorrelated factors, such as synoptic conditions of the eddy fields, effectively isolating the impact of baroclinicity on individual storm growth.

First, the connection between BC and the MI (Eq.~\ref{eq:intensity}) of the cyclones is examined (Fig.~\ref{fig:prop_corr}). For values of BC lower than 2.5~day$^{-1}$, the MI is positively correlated to BC (correlation of 0.96). This finding aligns with our expectation that regions of enhanced baroclinicity, such as the storm tracks, would be associated with increased storm activity. However, the positive correlation breaks at high BC. Indeed, calculating the correlation only for BC values larger than 2.5~day$^{-1}$ results in a correlation of -0.37. This non-linearity might play a significant role in the dynamics of regions characterized by extreme baroclinicity, such as the Pacific during midwinter \citep{Nakamura1992}. This explains why the cyclonic contribution to EKE saturates during midwinter \citep{Okajima2023} even though the track density peaks during midwinter \citep{Okajima2021}. For anticyclones, the MI increase rate slows down at high baroclinicity (0.95 and 0.80 correlation for values below and above 2.5~day$^{-1}$ respectively, Fig.~\ref{fig:prop_corr}d). 

Dividing storms into equatorward, central, and poleward jet regions (Fig.~\ref{fig:prop_corr}a; blue, orange, and green dots, respectively) reveals that cyclones are strongest on the poleward side of the jet and weakest on the equatorward side, while anticyclones exhibit the opposite trend.

To explain the non-monotonic behavior of the MI-BC relation, the MI is decomposed into GT and LGR, defined as the actual average rate of intensification during the growth stage, as measured by the tracking algorithm:
\begin{align}
\text{MI}_{\text{cyclones}} & \approx\text{LGR}\times\text{GT},   
\end{align}
where the equation is exact up to the initial intensity. 

The LGR monotonically increases with BC for all values (0.96 for both cyclones and anticyclones Fig.~\ref{fig:prop_corr}b,e). Since BC closely resembles the prediction for the baroclinic growth rate in linear instability models \citep[e.g.,][]{charney1947,Eady1949,Phillips1954}, its linear relationship with LGR suggests that these models accurately capture the average cyclone growth.

However, the GT is negatively correlated with BC (correlation of -0.96 and -0.92 for cyclones and anticyclones, respectively, Fig.~\ref{fig:prop_corr}c,f), indicating that the reduction in GT is responsible for the decrease in MI at high BC values. The reduction in GT with increasing baroclinicity aligns with previous meteorological studies of the West Pacific \citep{Schemm2021}, which showed that GT reduces from summer to winter, opposing the seasonal trend in baroclinicity. This trend was also observed in idealized GCM simulations, which demonstrated that it stemmed from the mean state’s influence on the vertical structure of baroclinic waves \citep{Hadas2021suppression}. 

Decomposing the contribution from different regions relative to the jet shows that cyclones on the poleward side of the jet have a higher LGR (Fig.~\ref{fig:prop_corr}b, blue versus green and orange), consistent with their larger MI. In addition, cyclones on the center of the jet have a higher GT (Fig.~\ref{fig:prop_corr}c, orange versus green and blue), which is consistent with their greater MI compared to cyclones on the equatorward side of the jet. Anticyclones on the poleward side of the jet have slightly higher LGR and much lower GT, such that their MI is lower (Fig.~\ref{fig:prop_corr}d-f, blue versus orange and green).

The observed LGR and GT trends indicate that the increase in MI at most BC values can be attributed to the increase in LGR, while the decline at high BC is driven by the decrease in GT. To quantify the inherent nonlinearity arising from the competing effects of LGR and GT, a linear model is applied to relate LGR and GT to BC (Fig.~\ref{fig:prop_corr}b,c,e,f solid, the coefficients for cyclones and anticyclones are given in Tab.~S1 and Tab.~S2, respectively). The constant coefficients for LGR are small (for the range of BC relevant to Earth), which fits our expectation that no baroclinicity will result in no growth. The main difference in the coefficients between cyclones and anticyclones is that the linear coefficient for the LGR is about three times larger for cyclones. In contrast, the fitted coefficients for the GT are remarkably similar. 

Using the fitted coefficients, the MI is approximated as the multiplication of the two curves, which results in a parabola. The resulting estimate for MI is given by:

\begin{align}
\text{MI}_{\text{cyclones}} & \approx12.0 \text{BC}-1.4 \text{BC}^{2}\label{eq:Gr cyclones},\\
\text{MI}_{\text{anticyclones}} & \approx4.4 \text{BC}-0.48 \text{BC}^{2},\label{eq:Gr anti}
\end{align}
where the constant term is neglected because it is small. The units of the linear and quadratic coefficients are hPa$\cdot$day and hPa$\cdot$day$^2$, respectively. The first term can be interpreted as the linear increase in MI due to the increase in LGR, while the second term can be interpreted as the nonlinear decrease in MI due to the decrease in GT. For low and medium values of BC (1~day$^{-1}$), the linear term contributes about nine times more than the nonlinear term, while for high BC (2.5~day$^{-1}$), it contributes only about three times more. This demonstrates why the linear approximation works well for most values of baroclinicity relevant to Earth's climate, but breaks down for high baroclinicity values.

These findings highlight the value of the Lagrangian perspective in studying mean-eddy interactions by resolving the complexities of individual storm growth. The results show that MI, which is expected to correlate with Eulerian EKE \citep{Schemm2018}, increases monotonically with BC at low and medium baroclinicity levels but decreases for cyclones and saturates for anticyclones at higher values. The LGR increases monotonically with baroclinicity, aligning with predictions from linear baroclinic instability models. However, the observed nonlinearity at high baroclinicity arises due to a reduction in GT. Additionally, we find a significant difference in the growth rates of different baroclinic wave phases, with cyclones exhibiting much faster growth. Finally, the location relative to the jet (which affects the barotropic shear) influences the growth of individual cyclones and anticyclones, a topic further explored in Sec.~\ref{subsec:dist}.

\subsection{Theoretical view on the empirical response \label{subsec:theory}}

To identify a potential physical mechanism underlying the nonlinear behavior of MI, the increase in LGR, and the reduction in GT with baroclinicity, this section constructs a minimal idealized model that reproduces the observed trends (Sec.~\ref{subsec:Correlations}). To build intuition, we first consider an overly simplified model. Following \cite{davies1994eady} and using linear theory, the time evolution of two counter-propagating Rossby waves with common amplitude $A$, and a relative phase of $\epsilon$ in a zonal flow with shear $\Lambda$, Brunt-Väisälä frequency $N$ and Coriolis parameter $f$ (characterized by BC$=\frac{\Lambda f}{N}$), is given by:

\begin{align}
    \frac{\partial{A}}{\partial{t}} &= - \text{BC}\sin{(\epsilon)}A\Delta,  \label{eq:A_t} \\
    \frac{\partial{\epsilon}}{\partial{t}} &= \text{BC}\left( \alpha -\cos{(\epsilon)} \right) 2\Delta, \label{eq:e_t}
\end{align}
where $\alpha$ is a geometric factor proportional to the ratio between the disturbance length scale and the Rossby radius of deformation (typically slightly smaller than 1 for cyclones), and $\Delta$ is a geometric factor related to zonal and meridional length scales (typically close to 1).

These equations indicate that wave growth requires a phase shift between 0 and $-\pi$ and that the rate of phase change depends on baroclinicity. Dividing both sides by BC allows reformulation in terms of normalized time ($\frac{\partial{}}{\partial{t\cdot\text{BC}}}$). From this perspective, higher baroclinicity shortens the timescale, accelerating both growth rate and phase change, which can reduce GT \citep{davies1994eady,Badger2001simple,Tamarin2015nonnormal}. 

Supporting this idea, \cite{Hadas2021suppression} used an aquaplanet GCM to show that shorter GT with increasing baroclinicity results from faster upper-level eddy drift, driven by a stronger jet stream. Similarly, \cite{schemm2019efficiency} found that meridional and zonal tilt contribute to storm suppression during the Pacific midwinter minimum. Thus, we hypothesize that the influence of baroclinicity on phase evolution drives the observed reduction in GT and the resulting nonlinear trend of MI in the empirical results.

However, the linear dynamics of these equations are not characterized by an MI or GT, as the normal mode solution has a constant phase ($\epsilon$) and indefinitely growing amplitude ($A$), a typical limitation of linear models. Therefore, a more complex model is required to test our hypothesis. To introduce greater complexity and establish a connection to the energetic perspective, we construct a framework based on the Lorenz energy cycle \citep[e.g.,][]{Lorenz1955,Oort1983}:

\begin{align}
    \frac{{\partial}P_e}{{\partial}t} &= \text{CP} + \underbrace{\frac{1}{\rho\overline{T}}\omega'T'}_\text{CE},\label{eq:LorenzP}\\
    \frac{{\partial}K_e}{{\partial}t} &= -\text{CK} - \underbrace{\frac{1}{\rho\overline{T}}\omega'T'}_\text{CE},\label{eq:LorenzK}
\end{align}
where $P_e$ and $K_e$ are the eddy available potential energy and kinetic energy, $\omega$ is the vertical velocity, CP and CK are the baroclinic and barotropic conversion (given by Eq.~\ref{Eq_CPEM} \& \ref{Eq_CKEM}, respectively). The last term in both equations is the conversion of $P_e$ to $K_e$ (CE). Assuming, for simplicity, that:

\begin{align}
    u',v' \sim \sqrt{K_e} ~,~ T',\omega'\sim \sqrt{P_e}~,
\end{align}
where $\omega$ is assumed to scale like the eddy potential energy as convection will be triggered by temperature anomalies created by baroclinic conversion at lower levels. Therefore, the eddy part of the conversion terms can be written as:

\begin{align}
        \text{CK}_\text{eddy} \sim K_e ~,~ \text{CP}_\text{eddy}\sim \sqrt{K_e{\cdot}P_e}~,~\text{CE}_\text{eddy}\sim P_e,
\end{align}
and then Eqs.~\ref{eq:LorenzP} \& \ref{eq:LorenzK} can be expressed as:

\begin{align}
    \frac{{\partial}P_e}{{\partial}t} &=\underbrace{b\sqrt{K_e{\cdot}P_e}}_\text{CP} - \underbrace{cP_e}_\text{CE}, \label{Eq_pe}  \\
    \frac{{\partial}K_e}{{\partial}t} &=-\underbrace{aK_e}_\text{CK} + \underbrace{cP_e}_\text{CE}, \label{Eq_ke}
\end{align}
where the coefficients $a,~b,~c$ have units of rate and incorporate in them two crucial properties: the correlation between the eddy terms and the dependence on the mean flow conditions. 

Next, the competing effects of baroclinicity on phase and amplitude evolution in the linear model (Eqs.~\ref{eq:A_t} \& \ref{eq:e_t}) is incorporated into the more complex model (Eqs.~\ref{eq:LorenzP} \& \ref{eq:LorenzK}). In the real atmosphere, storm growth depends on the phase between the upper and lower part of the storms, as westerly tilted anomalies create a positive correlation between lower-level temperature and wind perturbations, which is essential for baroclinic energy conversion from the mean flow to the eddies \citep[Eq.~\ref{Eq_CPEM},][]{Hoskins2014fluid}. Thus, the indirect effect of baroclinicity on storm growth—through its influence on phase evolution ($\epsilon$), can be incorporated in a simplified way by modifying the efficiency of baroclinic conversion as a function of time:

\begin{align}
    b(t) = \text{BC}(1-\text{BC}/\tau t), \label{eq:B_t}
\end{align}
where $\tau$ is a non-dimensional parameter that sets the relation between the BC and the rate of change in $\epsilon$. Comparing Eq.~\ref{eq:A_t} and Eq.~\ref{eq:B_t}, the time dependence can be interpreted as an upper-level wave that starts with an initial phase of  $-\frac{\pi}{2}$, and its phase evolves over time in such a way that the efficiency of baroclinic conversion decreases linearly with time.

\begin{figure}
    \centering
    \includegraphics[width=31pc]{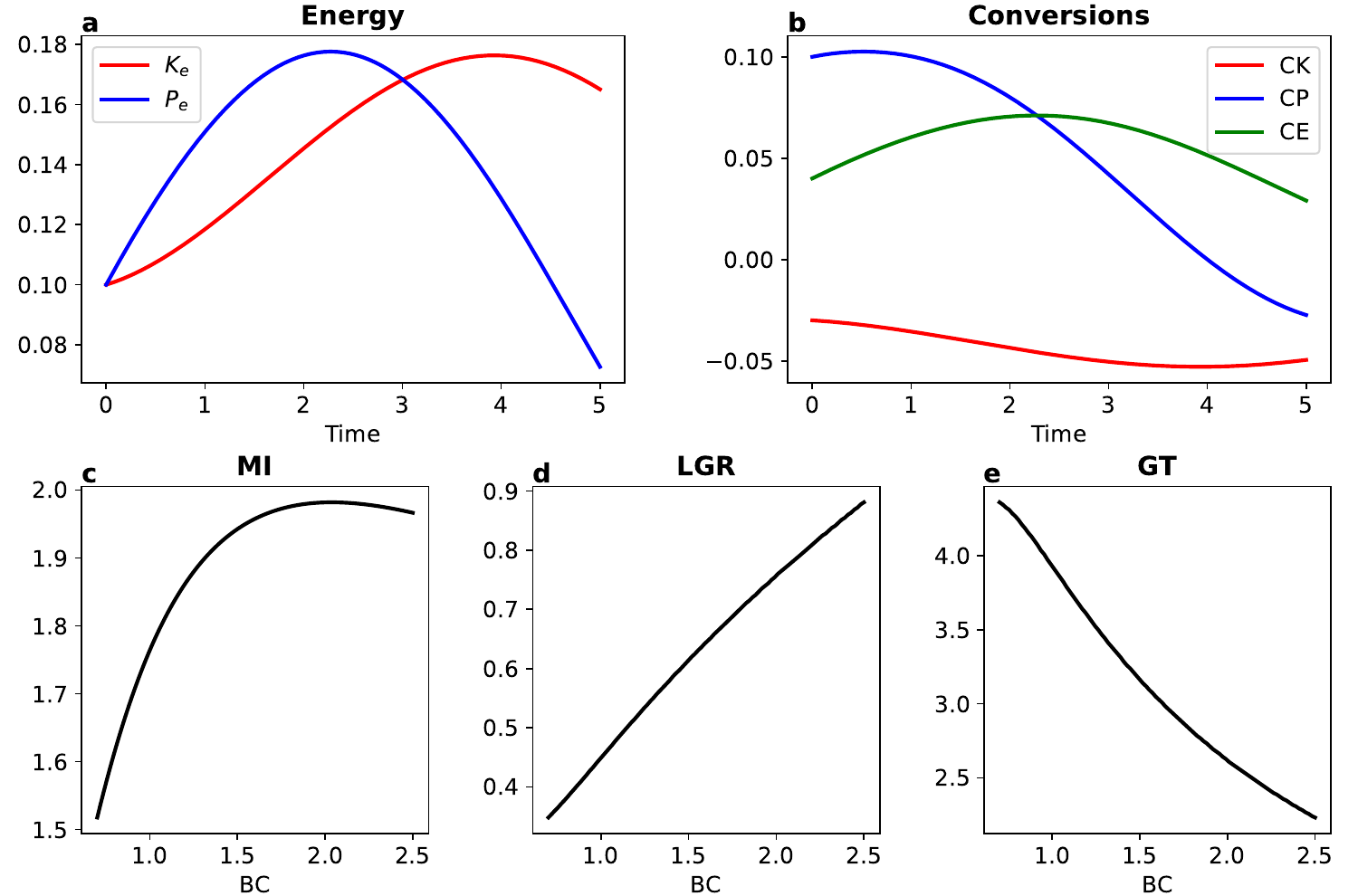}
    \caption{(a) The solution of Eq.~\ref{Eq_pe} and Eq.~\ref{Eq_ke} for $a=0.4~,~\text{BC}=1~,~c=0.4~,~\tau=4$ for initial pertubations $P_e(0)=0.1~,~K_e(0)=0.1$ as function of time. (b) The contribution of baroclinic conversion (CP, blue), the conversion of $K_e$ to $P_e$ (CE, green), and the barotropic conversion (CK, red) to the energy budget as a function of time. (c-e) The MI (defined as the maximum of $K_e$, c), the LGR (defined as the average rate of change of $K_e$, d) and the GT (defined as the time step of maximum $K_e$, e) as a function of BC, with $a$, $c$ and $\tau$ fixed to the values used in (a).}
    \label{fig:theory}
\end{figure}

To demonstrate the relevance of this equation to the idealized lifecycle of storms, Eq.~\ref{Eq_pe} and Eq.~\ref{Eq_ke} are integrated over time, starting from a westward tilted with height small anomaly. Using $a=0.4~,~\text{BC}=1~,~c=0.4~,~\tau=4$, results in a realistic idealized energy lifecycle (comparing for example to \cite{Simmons1978}): Initially, $P_e$ increases (Fig.~\ref{fig:theory}a blue) due to CP (Fig.~\ref{fig:theory}b blue). Then, as the phase decreases, CE surpasses CP (Fig.~\ref{fig:theory}b blue versus green), causing $P_e$ to decline while $K_e$ increases (Fig.~\ref{fig:theory}a blue and red). Finally, as $P_e$ reduces, the CK exceeds CE (Fig.~\ref{fig:theory}b red versus green), ending the $K_e$ growth. 

Sensitivity tests on BC reveal a trend similar to that in  Sec.~\ref{subsec:Correlations}. An analog to MI can be constructed as the maximum of $K_e$. Solving the equations for a range of BC values (holding $a,~c,~\tau$ constant) results in an increase in MI for small BC values and saturation or even decrease for high BC values (Fig.~\ref{fig:theory}c). To further analyze this behavior, MI is decomposed into GT and LGR, where LGR is defined as the average growth rate during the increase in $K_e$ and GT is the time at which $K_e$ reaches its maximum. This decomposition reveals that while LGR consistently increases across the entire BC range (Fig.~\ref{fig:theory}d), GT decreases (Fig.~\ref{fig:theory}e), aligns with the trends presented in Sec.~\ref{subsec:Correlations}.

The results demonstrate that an idealized model based on the Lorenz energy cycle successfully reproduces the observed trends of MI, LGR, and GT with baroclinicity (Sec.~\ref{subsec:Correlations}) when incorporating a coupling between baroclinicity and the rate of change in baroclinic conversion efficiency. This suggests that the shorter GT and nonlinear decrease in MI can be attributed to a faster decline in baroclinic conversion efficiency with increasing baroclinicity, driven by accelerated phase changes between upper- and lower-level waves. These phase changes are induced by the strong vertical wind shear characteristic of high baroclinicity. Future studies should extend the model to include nonlinear effects, such as phase locking of baroclinic waves and its disruption by cascades and interactions with shorter waves \citep{Tamarin2015nonnormal} and diabatic effects, such as latent cooling and surface fluxes \citep{Haualand2019does,Haualand2020direct}, which might explain the deviations of the empirically found trends from the results of the idealized model.

\section{The Lagrangian response to the full atmospheric mean state\label{subsec:dist}}

Although barotropic properties of the mean flow are known to have a profound influence on the dynamics of storms \citep{Lorenz1955,James1987,Riviere2013,Chemke2022}, which is supported by the difference in growth between storms in different locations relative to the jet (Fig.~\ref{fig:prop_corr}), they are rarely considered when assessing the storm track response to climatic forcing. Therefore, in this section, the influence of the barotropic shear of the mean flow on storm growth is systematically studied, and compared to the influence of the baroclinic properties. 

The shear rates (Sec.~\ref{subsec:Energetic-Perspective}) quantify the barotropic and baroclinic properties of the mean flow. Barotropic properties are measured using $\widehat{\text{CK}}_{x}$, and $\widehat{\text{CK}}_{y}$, which are proportional to horizontal shear (Eq.~\ref{eq:CK}), while baroclinic properties are quantified by $\widehat{\text{CP}}_{x}$, and $\widehat{\text{CP}}_{y}$, which are proportional to vertical shear (Eq.~\ref{eq:CP}). The response of storms' growth to each shear component is analyzed by sorting cyclones and anticyclones based on the composite and temporal average shear rates during their growth stage and examining the resulting differences in LGR and GT (Sec.~\ref{subsec:Storm-tracking}). Additionally, a 4-dimensional linear model is fitted to quantify the relative importance of each shear rate term (see Sec.~S1).

The discussion begins by examining the frequency at which storms experience different conversion rates. It then explores the influence of shear rates on LGR, which is identified as the most complex aspect, following a structured approach. First, the effect of baroclinic characteristics on cyclones is analyzed, followed by an investigation of the barotropic characteristics. This is then extended to assess the relative importance of baroclinic versus barotropic properties before comparing the differences between cyclones and anticyclones. After addressing these aspects, the impact on GT is analyzed. Finally, the findings are synthesized into a comprehensive summary of how baroclinic and barotropic properties influence storm growth.

\begin{figure}
\begin{centering}
\includegraphics[width=31pc]{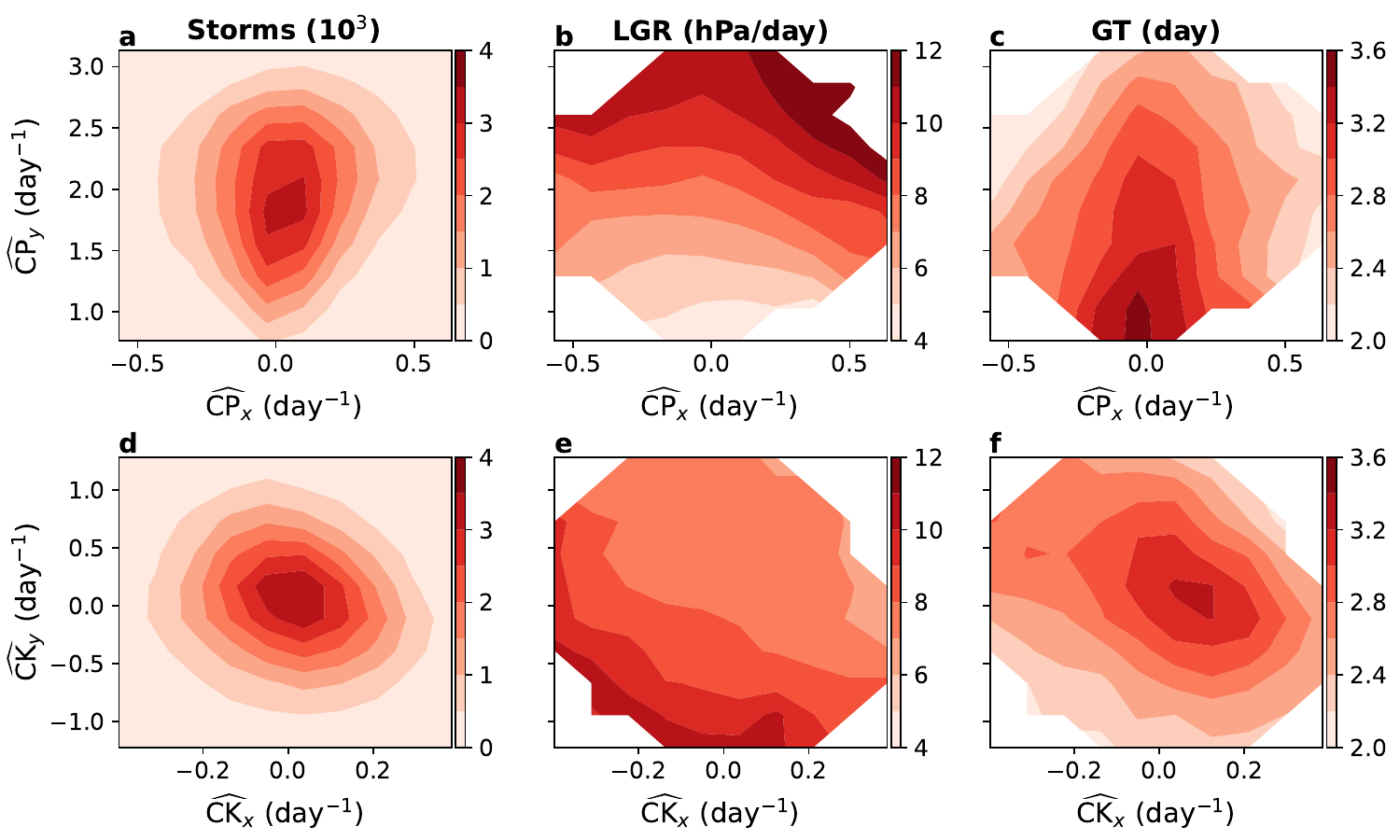}
\par\end{centering}
\caption{\label{fig:2dist} Number of storms (a,d), LGR (hPa$\cdot$day$^{-1}$,
b,e) and GT (days, c,f) as a function of $\widehat{\text{CP}}_{x}$
and $\widehat{\text{CP}}_{y}$ (day$^{-1}$, a-c) and $\widehat{\text{CK}}_{x}$
and $\widehat{\text{CK}}_{y}$ (day$^{-1}$, d-f).}
\end{figure}

First, the distribution of cyclones and anticyclones in the shear rate space ($\widehat{\text{CP}}_{x}$, $\widehat{\text{CP}}_{y}$, $\widehat{\text{CK}}_{x}$, $\widehat{\text{CK}}_{y}$) is analyzed. The shear rate coefficients generally exhibit low correlation (Tab.~S3 and Tab.~S4) and are therefore treated as independent variables.
Storms typically experience positive values of $\widehat{\text{CP}}_{y}$, indicating easterly jets, which are characteristic of storm development regions \citep[e.g.,][]{Chang2002}. The other coefficients are centered around zero, with both positive and negative values observed. The key distributional difference between cyclones and anticyclones is that anticyclones experience higher values of $\widehat{\text{CK}}_{y}$, meaning they are more frequently located on the anticyclonic (equatorward) side of the jet. This aligns with previous findings showing that anticyclone track density is more equatorward \citep{hoskins2002,hoskins2005}.

 The effect of the baroclinic terms on cyclones' LGR (Fig.~\ref{fig:2dist}b) shows a positive correlation with $\widehat{\text{CP}}_{y}$ and the absolute value of $\widehat{\text{CP}}_{x}$. This result aligns with findings from Sec.~\ref{subsec:Correlations} and previous studies showing a strong connection between high growth rates and high baroclinicity \citep[e.g.,][]{hoskins2002}. A linear model indicates that, given the fitted coefficients and the relevant value range, $\widehat{\text{CP}}_{y}$ has three times the impact of $\widehat{\text{CP}}_{x}$. 
 
 Cyclones' LGR is negatively correlated with both barotropic shear rates (Fig.~\ref{fig:2dist}e). Linear Regression analysis shows that $\widehat{\text{CK}}_{x}$ and $\widehat{\text{CK}}_{y}$ have similar contribution. The negative correlation between $\widehat{\text{CK}}_{y}$ and LGR fits the results in Fig.~\ref{fig:prop_corr}b, as cyclones on the poleward side of the jet exhibit lower $\widehat{\text{CK}}_{y}$ and higher LGR, while the opposite holds for cyclones on the equatorward side. Furthermore, the high LGR at negative values of both $\widehat{\text{CK}}_{y}$ and $\widehat{\text{CK}}_{x}$ is consistent with expectations that storms on the poleward side of the jet exit region experience enhanced cyclonic growth, driven by ageostrophic wind divergence at high levels \citep{Martin2013mid}. 

\begin{figure} 
\begin{centering} 
\includegraphics[width=31pc]{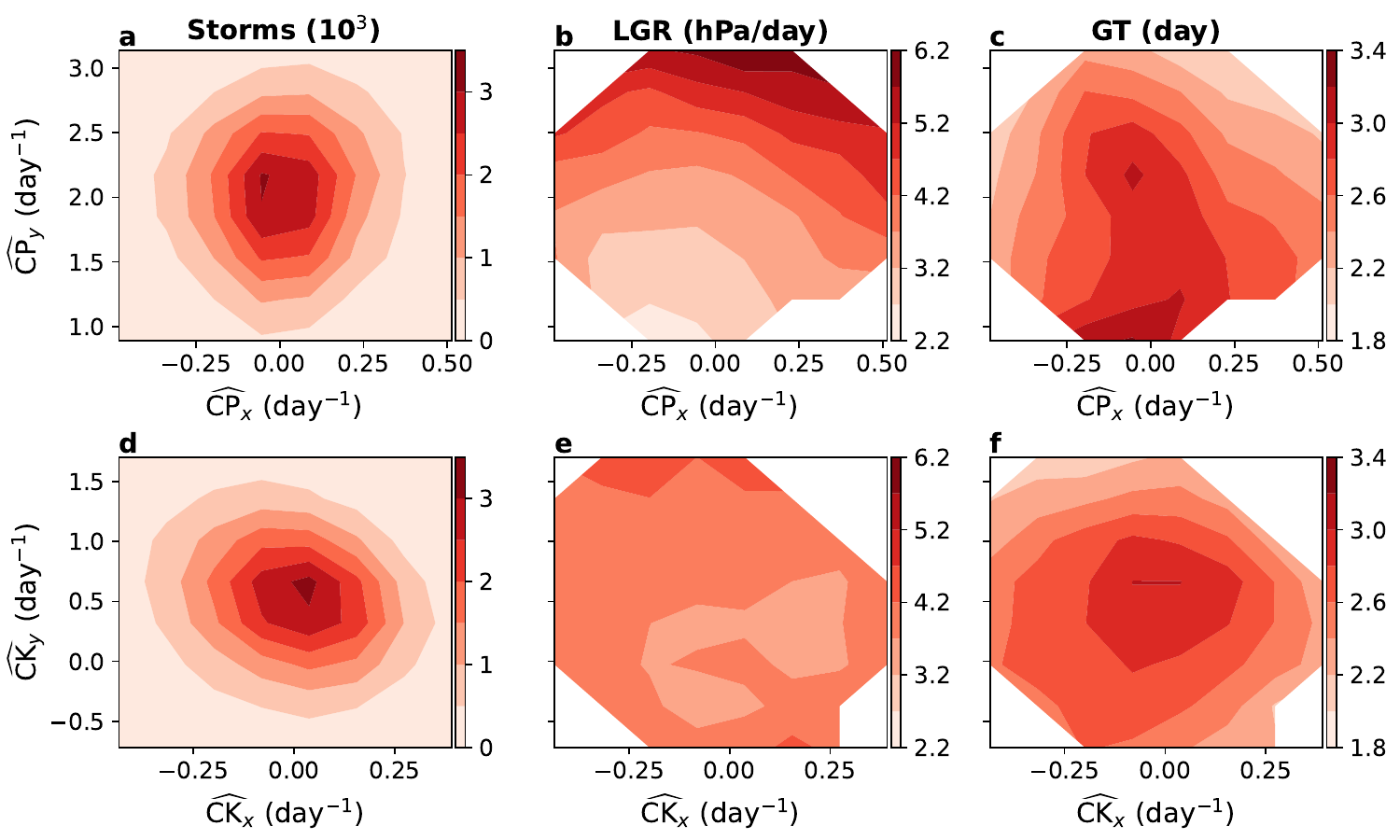} \par\end{centering} \caption{\label{fig:2dist_acyc}Same as Fig.~\ref{fig:2dist}, but for anticyclones.} 
\end{figure}

Comparing the contribution of the barotropic and baroclinic shear rates to cyclones' LGR (Fig.~\ref{fig:2dist}b versus Fig.~\ref{fig:2dist}e) reveals that the baroclinic contribution is approximately twice as large. For anticyclones, the barotropic influence on LGR is small, being about four times smaller than the baroclinic contribution (Fig.~\ref{fig:2dist_acyc}b versus Fig.~\ref{fig:2dist_acyc}e). Additionally, linear analysis indicates tha $\widehat{\text{CP}}_{y}$s roughly twice as influential as $\widehat{\text{CP}}_{x}$. 

The GT of both cyclones and anticyclones responds similarly to the barotropic and baroclinic properties of the mean flow, decreasing with increasing shear rates (Fig.~\ref{fig:2dist}c, Fig.~\ref{fig:2dist}f, Fig.~\ref{fig:2dist_acyc}c, and Fig.~\ref{fig:2dist_acyc}f), at a comparable rate. Additionally, all four conversion rates appear to influence GT similarly. This pattern aligns with the observation that cyclones both poleward and equatorward of the jet exhibit shorter GT than those near the jet core (Fig.~\ref{fig:prop_corr}c orange vs green and blue). 

A key difference between cyclones and anticyclones is the optimal $\widehat{\text{CK}}_{y}$ for GT, which is about 0.5~day$^{-1}$ for anticyclones, compared to about 0.1 for cyclones (Fig.~\ref{fig:2dist}f vs Fig.~\ref{fig:2dist_acyc}f).  This aligns with the observation that poleward anticyclones have much lower GT, while those near the jet core and equatorward of the jet exhibit similar GT values (Fig.~\ref{fig:prop_corr}f orange vs green). Since very high values of $\widehat{\text{CK}}_{y}$ are rare (Fig.~\ref{fig:2dist_acyc}d), this may explain the observed distribution. A possible explanation for the longer GT of equatorward cyclones and anticyclones is that some storms in this region may initially form as thermal lows, which later transition into baroclinic storms \citep{Campins2000catalogue}, extending their overall GT.

Building on Sec.~\ref{subsec:Correlations}, this analysis reveals key details about the mean flow’s impact on storm growth. For LGR, baroclinic properties play a larger role for cyclones and dominate for anticyclones, while the mean zonal flow has a stronger influence than the meridional flow. In contrast, GT is equally affected by baroclinic and barotropic properties, as well as by zonal and meridional winds. An extension of the mechanism proposed in Sec.~\ref{subsec:theory} could be suggested to explain this trend: the horizontal and vertical shear of the flow disrupt the baroclinic structure necessary for growth through differential advection. Consequently, stronger shear, characterized by higher baroclinic and barotropic shear rates, leads to a reduction in the growth time of the wave. This generalization of the mechanism fits previous results of \cite{schemm2019efficiency}, which showed over the Pacific during midwinter that meridional tilt has a similar effect on baroclinic conversion to zonal shear. Further research, which includes diabatic effects, is needed to fully understand the underlying mechanism.

\section{Summary}\label{sec:summary}

When investigating the midlatitude climate's response to changes in atmospheric forcing, whether natural or anthropogenic, a key challenge lies in understanding how eddies respond to the mean state of the atmosphere. While many studies have explored this interaction from an Eulerian perspective, there has yet to be a systematic exploration of how the mean state of the atmosphere impacts the growth of storms from a Lagrangian perspective. To address this gap, this study investigates the statistical relationship between commonly used baroclinic and barotropic properties of the mean flow and the growth of individual cyclones and anticyclones.

A systematic quantification of the influence of baroclinicity (Eq.~\ref{eq:EGR}), on Maximum Intensity (MI, Fig.~\ref{fig:prop_corr}a,d) shows that MI generally increases with baroclinicity. However, at extreme baroclinicity values, this trend reverses for cyclones and significantly slows down for anticyclones. This behavior is particularly relevant to midlatitude weather responses to extreme baroclinicity, such as those observed over the North Pacific during midwinter.

To explain this nonlinearity, MI is decomposed into Lagrangian Growth Rate (LGR) and Growth Time (GT). LGR increases monotonically with baroclinicity (Fig.~\ref{fig:prop_corr}b,e), confirming that the growth rate predicted by linear baroclinic instability models aligns with observed Lagrangian growth. However, GT decreases with baroclinicity (Fig.~\ref{fig:prop_corr}c,f), indicating that the linear increase in MI at most baroclinicity values is driven by the rapid rise in LGR, while the MI decline at extreme baroclinicity for cyclones results from nonlinear effects due to GT reduction. Based on these empirical results, the magnitudes of the linear and nonlinear terms are estimated, leading to a nonlinear correction to the classical relationship between baroclinicity and storms' activity (Eqs.~\ref{eq:Gr cyclones} \& \ref{eq:Gr anti}).

The observed relationship between baroclinicity, MI, LGR, and GT is successfully reproduced (Fig.~\ref{fig:theory}) using a simple growth model based on the Lorenz energy cycle, where the rate of change in baroclinic conversion efficiency is coupled to baroclinicity (Eqs.~\ref{Eq_ke}-\ref{eq:B_t}). This demonstrates that the reduction in GT and the nonlinear decrease in MI can arise from a faster decline in baroclinic conversion efficiency with increasing baroclinicity, which is intuitively linked to phase shifts between the upper- and lower-level waves. In reality, this coupling results from faster differential advection of the upper-level wave relative to the lower-level wave, driven by enhanced vertical shear of the mean flow under higher baroclinicity conditions.

Incorporating the barotropic properties of the mean flow (horizontal shear) into the analysis reveals that they have a weaker influence on LGR, being twice less influential for cyclones and four times less influential for anticyclones (Fig.~\ref{fig:2dist}b vs. Fig.~\ref{fig:2dist}e and Fig.~\ref{fig:2dist_acyc}b vs. Fig.~\ref{fig:2dist_acyc}e). However, GT for both cyclones and anticyclones is strongly affected by both barotropic and baroclinic characteristics of the mean flow (Fig.~\ref{fig:2dist}c,f and Fig.~\ref{fig:2dist_acyc}c,f). 

The complex dynamics revealed by this study underscore the importance of considering the response of individual storms to climatic changes. The findings demonstrate that the commonly assumed linear relationship between storm activity and mean baroclinicity holds only at low baroclinicity levels, highlighting the need for caution when applying it to high-baroclinicity regions. However, they also show that idealized dry models effectively capture key aspects of the observed trends. Furthermore, the results underscore the importance of accounting for both barotropic and baroclinic properties of the mean flow when using eddy–mean flow frameworks to study storm track responses to climatic conditions.

These findings raise several fundamental open questions. A key unresolved issue is the relative importance of mean flow versus eddy flow in individual storm growth, which is critical for the assessment of storms' internal variability for attribution efforts. Another crucial question is the physical mechanism underlying the observed trends, particularly the reduction in GT with barotropic properties, and the role of processes such as diabatic heating in this relationship. Additionally, this study focuses on how the mean flow influences storms, but the reverse interaction—how storms feedback onto the mean flow—remains unexplored. Future research should address this feedback to develop a comprehensive understanding of eddy-mean flow interactions.

\section{Open Research}
\subsection{Data Availability Statement}

No new data sets were generated during the current study. ERA-5 is available through the Climate Data Store (\url{cds.climate.copernicus.eu}).

\acknowledgments
This research has been supported by the Azrieli fellowship and the Israeli Science Foundation (Grant 996/20).

\bibliography{orsbib.bib}

\appendix
\section{Baroclinic and barotropic energy conversion}\label{sec:energy_der}

To obtain the equation for barotropic conversion, Eq.~4 in \cite{Simmons1983barotropic} is multiplied by $rho$ to translate into units of energy per volume:
\begin{align}
    C(K_{e},K_{m}) & =\rho\left(\overline{v'^{2}-u'^{2}},-\overline{u'v'}\right)\cdot
    \left(
    \frac{\partial\overline{u}}{\partial x}-\overline{v}\frac{\tan\phi}{r_{e}},
    \cos{\phi}\frac{\partial}{\partial y}\left(\frac{\overline{u}}{\cos{\phi}}\right)+\frac{\partial\overline{v}}{\partial x}\right)\underbrace{\rightarrow}_{\frac{\partial}{\partial y}=\frac{1}{r_e}\frac{\partial}{\partial \phi}} \\
    & =
    \rho\left(\overline{v'^{2}-u'^{2}},-\overline{u'v'}\right)\cdot\left(\frac{\partial\overline{u}}{\partial x}-\overline{v}\frac{\tan\phi}{r_{e}},\frac{\partial\overline{u}}{\partial y}+\frac{\partial\overline{v}}{\partial x}+\overline{u}\frac{\tan\phi}{r_{e}}\right).
\end{align}

To obtain the equation for baroclinic conversion, we multiply the second term in the right-hand-side of Eq.~4.2 in \cite{Orlanski1991} with $\rho$:
\begin{align}
    C(P_{e},P_{m}) & =\frac{1}{ \overline{\theta} \frac{\partial \overline{\theta}}{\partial p}}\left(\overline{u'\theta'},\overline{v'\theta'}\right)\cdot
    \left(\frac{\partial\overline{\theta}}{\partial x},\frac{\partial\overline{\theta}}{\partial y}\right).
\end{align}
Next, thermal wind balance is used to translate the gradient of potential temperature to wind shear. Furthermore, we cancel the Exner function between the $\overline{\theta}$ and $\theta'$ terms in the left part of the dot product. Finally, we express the mean potential temperature vertical gradient in terms of the Brunt–Väisälä frequency ($N$) using Eq.~\ref{eq:bvf}. This leads to the final version presented in the main text:
\begin{align}
    C(P_{e},P_{m}) 
    & = \frac{\rho g}{\overline{N}\overline{T}}\left(\overline{u'T'},\overline{v'T'}\right)\cdot\frac{f}{\overline{N}}
    \left(\frac{\partial\overline{v}}{\partial z},\frac{\partial\overline{u}}{\partial z}\right).\label{eq:interCP}
\end{align}
\end{document}